# Experimental demonstration of asymmetric diffraction based on a passive parity-time-symmetric acoustic grating

Yuzhen Yang,[1, 2]   Han Jia,[1, 2, 3, *]   Yafeng Bi,[1, 2]   Han Zhao,[1, 2]   Jun Yang[1, 2, 3, †]

[1] Key Laboratory of Noise and Vibration Research, Institute of Acoustics, Chinese Academy of Sciences, Beijing 100190, People's Republic of China

[2] University of Chinese Academy of Sciences, Beijing 100049, People's Republic of China

[3] State Key Laboratory of Acoustics, Institute of Acoustics, Chinese Academy of Sciences, Beijing 100190, People's Republic of China

Passive parity-time-symmetric medium provides a feasible scheme to investigate non-Hermitian systems experimentally. Here, we design a passive *PT*-symmetric acoustic grating with a period equal to exact *PT*-symmetric medium. This treatment enhances the diffraction ability of a passive *PT*-symmetric grating with more compact modulation. Above all, it eliminates the first-order disturbance of previous design in diffraction grating. Additional cavities and small leaked holes on top plate in a 2D waveguide are used to construct a parity-time-symmetric potential. The combining between additional cavities and leaked holes makes it possible to modulate the real and imaginary parts of refractive index simultaneously. When the real and imaginary parts of refractive index are balanced in modulation, asymmetric diffraction can be observed between a pair of oblique incident waves. This demonstration provides a feasible way to construct passive parity-time-symmetric acoustic medium. It opens new possibilities for further investigation of acoustic wave control in non-Hermitian systems.

[*]Corresponding author.

hjia@mail.ioa.ac.cn

[†]Corresponding author.

jyang@mail.ioa.ac.cn

Parity-time (*PT*) symmetry is a concept arising from Quantum mechanics, which means the invariance of Hamiltonian under parity-time inversion. It was believed that Hermitian Hamiltonian was the guarantee of real spectra corresponding to observable physical quantities. Until 1998, Bender and Boettcher confirmed that non-Hermitian Hamiltonian could still possess real spectra in a complex system with parity-time reflection symmetry [1]. A necessary but not sufficient condition of *PT* symmetry is $V(\mathbf{r}) = V^*(-\mathbf{r})$. It means that the real part of the Hamiltonian potential is an even function of position, whereas the imaginary part is an odd function. When the complex potential exceeds a threshold, spontaneous breaking can be observed from the unbroken phase to broken phase of *PT* symmetry [2-5]. The phase transition singularity is called exceptional point, which is one of the most important characteristic of non-Hermitian Hamiltonian.

In recent years, non-Hermitian system with *PT*-symmetric potential has been widely applied in fields of classic waves because of the similarity between Schrodinger equation and classic wave equation [5-30]. Many unconventional phenomena in optics have been demonstrated, including Bloch oscillation [6, 7], optical isolation [8, 9], unidirectional invisibility or reflectionless effect [10-12], coherent perfect absorption (CPA) and lasing [15-20, 31]. In addition, many interesting acoustic effects of *PT* symmetry, such as one-way cloak [32], invisible sensing [22] and sound absorption [33], have also drawn much attention. However, previous works mainly pay attention to *PT* symmetry of 1D waveguides, in which wave propagates along with the direction of modulation. To enhance the practicability and applicability of *PT*-symmetric system, research on higher dimensional system is an urgent and important task [34-37]. Diffraction grating is a basic device in 2D waveguide, whose physical characteristics of *PT* symmetry have been discussed in optics [35-37]. Conventional Bragg scattering gratings with only real-part modulation of refractive index offer a pair of opposite wave vectors $\mathbf{q}' = \pm 2k_B \mathbf{r}/r$, where $k_B = \pi/T$ is the wave number on the Bragg condition. However, *PT*-symmetric gratings offer a unidirectional wave vector at EP, with balanced modulations between real and imaginary parts ($n_r = n_i$). In this letter, asymmetric diffraction of a passive *PT*-symmetric acoustic grating has been investigated theoretically and experimentally. At balanced modulation of real and imaginary parts, the +1st diffraction order with negative incident Bragg angle is much weaker than -1st diffraction order with positive incident angle. Such asymmetric diffraction acoustic grating have potential applications in many fields, such as beam forming, ultrasonic medical imaging, and directional noise reduction.

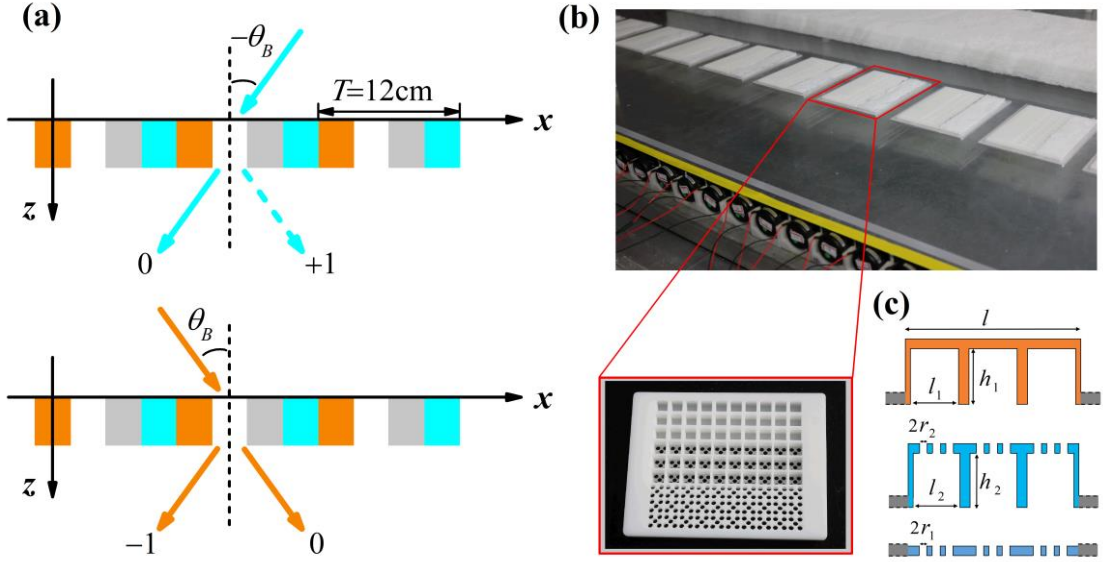

FIG. 1. An asymmetric diffraction grating based on *PT*-symmetric potential. (a) Upper figure shows diffraction with negative incident angle and lower figure shows diffraction with positive incident angle. The arrows represent indent waves and corresponding diffracted waves. (b) A photo of installation of sample pieces; the picture in red box is the photo of a sample piece fabricated by 3D printing technique. Each sample contains three kinds of modulators side by side. (c) Schematics of three kinds of modulators. The cross-sectional views of modulators for real, superposition of real and imaginary, imaginary parts are displayed from top to bottom. Adopted geometry parameters are $l = 30mm$, $l_1 = 7.6mm$, $h_1 = h_2 = 9.0mm$, $r_1 = 1.7mm$, $l_2 = 8.6mm$, and $r_2 = 1.5mm$, respectively.

In acoustics, *PT*-symmetric medium can be realized by a complex refractive index obeying the condition $n(\mathbf{r}) = n^*(-\mathbf{r})$. The classic *PT*-symmetric medium has simple harmonic modulation of refractive index $n(x) = n_0 + n_r \cos(hx) + i n_i \sin(hx)$, where $h = 2\pi/T$ is the reciprocal lattice vector, T is the lattice period, $n_0$ is the background refractive index and $n_r, n_i$ are the modulation amplitudes of the real and imaginary parts, respectively. The linear acoustic wave equation in the *PT*-symmetric grating can be written as

$$\Delta P + k_0^2 n^2 P = 0, \tag{1}$$

where $\Delta = \partial^2/\partial x^2 + \partial^2/\partial z^2$ is the Laplacian, P is the sound pressure, $k_0 = \omega/c_0$ is the wave number of background medium. The modulation part of the refractive index is given as $n_1(x) = n(x) - n_0$. In a weak coupling regime, the acoustic wave equation can be approximate to

$$\Delta P + k_0^2 (n_0^2 + 2 n_0 n_1) P = 0. \tag{2}$$

Here, we focus on the first-order Bragg scattering of the *PT*-symmetric grating. The high-order diffractions can be negligible in the studied frequency range. The Fourier expansion of the modulated refractive index is $n_1 = C_0 + C_{-1} \exp(ihx) + C_1 \exp(-ihx)$. For the incident monochromatic plane wave at Bragg angle $\theta_B$ [$2k \sin(\theta_B) = h$], as shown in Fig. 1(a), the sound pressure in the grating contains two main coupled diffractive waves. A general expression is written as

$$P(\mathbf{r}) = A_0 e^{-i\mathbf{q}_0 \mathbf{r}} + A_h e^{-i\mathbf{q}_h \mathbf{r}}, \tag{3}$$

where $A_0, A_h$ are amplitudes of the zero diffraction order and +1st (-1st) diffraction order, respectively; $\mathbf{q}_0 = (q_{0x}, q_{0z})$, $\mathbf{q}_h = (q_{hx}, q_{hz})$ are wave vectors of the zero order and +1st (-1st) order diffracted waves in the structure. When the sign of Bragg incident angle is negative, i.e. $-\theta_B$, the relationship between the components of wave vectors can be expressed as $q_{hx} = q_{0x} - h$, $q_{hz} = q_{0z}$. By substituting Eq. (3) into Eq. (2), we obtain the following coupled-mode equations

$$\begin{cases}(-q_{0x}^2 - q_{0z}^2 + k_0^2 n_0^2 + 2k_0^2 n_0 C_0)A_0 + 2k_0^2 n_0 C_{-1}A_h = 0 \\ 2k_0^2 n_0 C_1 A_0 + (-q_{hx}^2 - q_{hz}^2 + k_0^2 n_0^2 + 2k_0^2 n_0 C_0)A_h = 0\end{cases}, \quad (4)$$

where $q_{0x} = k_x$ [ $k_x = \sin(\theta_B)k_0$ ], which follows from Snell's law. The existence of nontrivial solutions of the system expressed by Eq. (4) allows us to write down the dispersion relation for $z$ projections of wave vectors

$$q_{0z}^2 = k_0^2 n_0^2 + 2k_0^2 n_0 C_0 - q_{0x}^2 \pm 2k_0^2 n_0 \sqrt{C_1 C_{-1}}. \quad (5)$$

The two modes of Eq. (5) represent upper and lower deviation of z-directional wave vector due to the periodical modulation of refractive index. Combining Eq. (4) and Eq. (5), the following relationship for the amplitudes of diffractive waves can be obtained:

$$R_{+1} = \frac{A_h}{A_0} = \pm\sqrt{\frac{C_1}{C_{-1}}}, \quad (6)$$

where $R_{+1}$ is the ratio between amplitude coefficients of +1$^{st}$ diffraction order and zero diffraction order. On the other hand, when the sign of the Bragg incident angle changes to positive, i.e. $\theta_B$, relationship between the x-directional projections of wave vectors in the *PT*-symmetric diffraction grating changes from $q_{hx} = q_{0x} - h$ to $q_{hx} = q_{0x} + h$. As a result, the Fourier coefficients in Eq. (4) exchange between $C_1$ and $C_{-1}$. The corresponding diffraction intensity ratio changes to $R_{-1} = A_h / A_0 = \pm\sqrt{C_{-1}/C_1}$ naturally. It can be seen that the diffraction intensity ratios of positive and negative incident angles are both determined by the Fourier coefficients of refractive index. In classic *PT*-symmetric system, the Fourier coefficients are shown as:

$$C_0^1 = 0, \quad C_1^1 = (n_r - n_i)/2, \quad C_{-1}^1 = (n_r + n_i)/2. \quad (7)$$

It is clear that $R_{+1}$ will equal to zero with balanced modulation between real and imaginary parts of refractive index, i.e. $n_r = n_i$, which means the +1$^{st}$ diffraction with negative incident angle will vanish completely at EP.

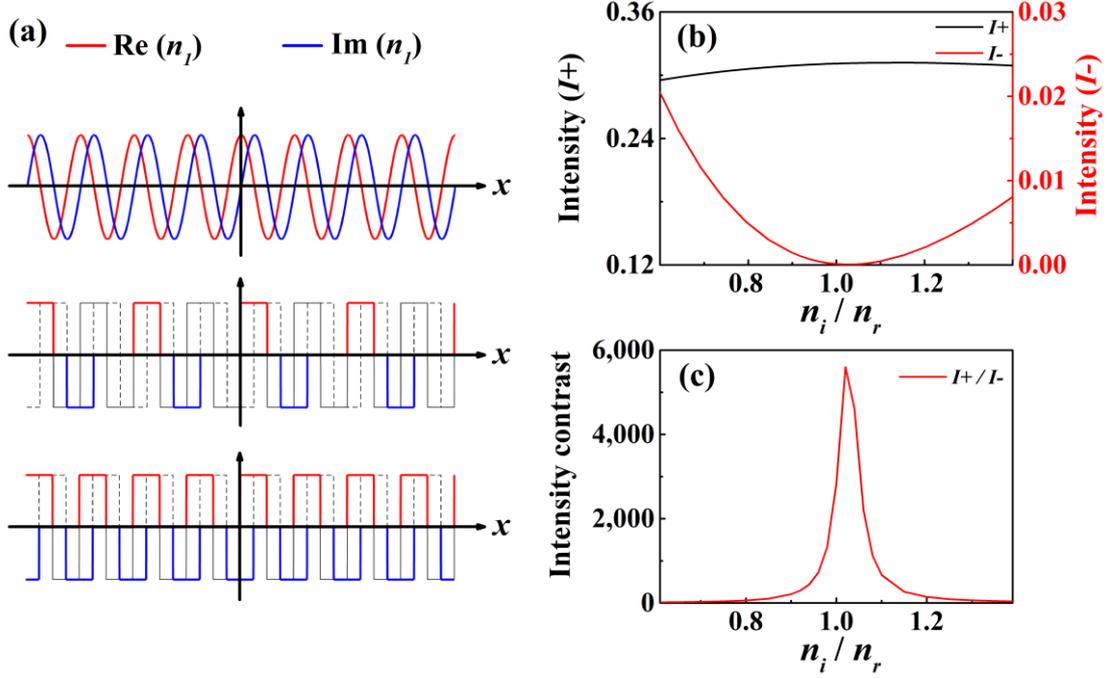

FIG. 2. Evolution of PT-symmetric potential and diffraction characteristic of the revised passive PT-symmetric potential. (a) The modulation of refractive index for exact PT-symmetric medium and passive PT-symmetric mediums. The red (blue) curves denote real (imaginary) part of modulation. The first row represents the modulation of exact PT-symmetric potential. The second row represents the existing popular passive PT-symmetric potential with doubled period. And the third row represents the revised passive PT potential with unchanged period. (b) Diffraction intensity with different modulation ratios. (c) Diffraction intensity ratio between -1$^{st}$ diffraction order and +1$^{st}$ diffraction order with different modulation ratios.

To simplify the structural design requirement, previous researchers adopted the Fourier translation of a complex square-wave modulation in exchange for the complex exponential modulation. In addition, the manual truncation and in-phase shift of refractive index separate the modulations of real and imaginary parts spatially. Considering the absence of natural gain medium, the gain part is removed to enhance the feasibility of experimental implementations. This evolution process from exact PT-symmetric modulation to passive PT-symmetric modulation is presented in Fig. 2(a). It has been proved that the features at EP of exact PT symmetry can still be observed and controlled in passive PT-symmetric system [11, 34, 37, 38]. However, this adjustment artificially doubles the modulation period. And the utilization of second-order Bragg scattering will leave unwished first-order disturbance in *PT*-symmetric diffraction gratings. Moreover, the expansion of the period reduces the ability of diffraction. To overcome the shortage of existing passive *PT*-symmetric medium in diffraction gratings, we change the period back to that of exact *PT*-symmetric medium, with overlap of real and imaginary modulations. The Fourier coefficients of refractive indexes in the existing and revised passive *PT*-symmetric mediums are shown as follows:

$$C_0^2 = \frac{1}{4}(n_r - in_i), \quad C_1^2 = \frac{1}{2\pi}(n_r - n_i), \quad C_{-1}^2 = \frac{1}{2\pi}(n_r + n_i), \tag{8a}$$

$$C_0^3 = \frac{1}{2}(n_r - in_i), \quad C_1^3 = \frac{1}{\pi}(n_r - n_i), \quad C_{-1}^3 = \frac{1}{\pi}(n_r + n_i). \tag{8b}$$

Comparing Fourier coefficients in Eqs. (8) to those of exact *PT* symmetry in Eq. (7), $C_0$ is not zero any more for passive *PT*-symmetric medium due to the absence of gain part. In passive *PT*-symmetric systems composed by only lossy and lossless parts, the average loss bias can be regarded as the loss of background medium. For both passive *PT*-symmetric mediums, $C_1, C_{-1}$ still have components of $n_r - n_i$ and $n_r + n_i$, respectively. So asymmetric diffraction effect still keeps valid in these passive *PT*-symmetric gratings. Moreover, the Fourier coefficients of revised grating are twice of the coefficients of existing *PT*-symmetric grating, which improve the scattering ability of grating obviously. Above all, the revised Bragg diffraction grating remains no disturbance of other diffraction orders, because the mutual coupling between two main waves in the grating is caused by the first-order Bragg scattering (see Section I of Supplement Material [39]). Diffraction in a revised grating with ideal parameters has been simulated with Comsol Multiphysics. The simulated results are shown in Figs. 2(b) and 2(c). I+ (I-) is the sound energy of -1st (+1st) diffraction order with positive (negative) incident angle. From Fig. 2(b), it can be seen that the +1st diffraction order is suppressed completely at EP, while -1st diffraction order is not very susceptible to the modulation ratio between real and imaginary parts. It is because unidirectional wave vector is generated in the diffraction grating when the modulation reaches the balanced condition

$n_r = n_i$. As long as the modulation deviates from the balanced condition, unidirectional mode coupling will be broken down and +1st diffraction order will appear again. Figure 2(c) shows the intensity contrast between -1st diffraction order and +1st diffraction order with different modulation ratios, which reaches a prominent peak at EP.

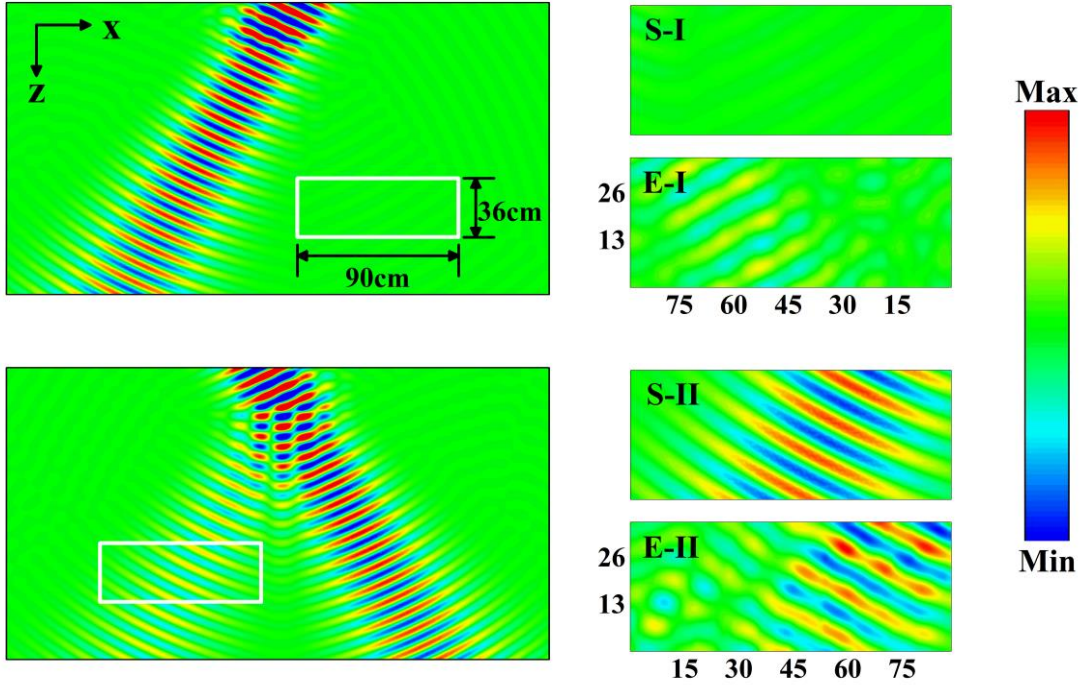

FIG. 3. Asymmetric Bragg diffraction of the revised *PT*-symmetric acoustic grating at 3.1 kHz. Left column shows the simulated sound pressure fields in the whole 2D waveguide. The +1st diffraction order is suppressed completely (upper figure) while the -1st diffraction order can be observed obviously (lower figure). Corresponding enlarged views of simulated (S-I and S-II) and measured (E-I and E-II) sound pressure fields are shown in right column. Remarkable contrast of different incident angles can be observed in both simulated and measured results.

To verify the asymmetric diffraction effect of the revised passive *PT*-symmetric grating, we designed an acoustic diffraction grating in a 2D waveguide. In the proposed model, three equal-width parts in a period need to be modulated. The cross-sectional views of corresponding modulators are shown in Fig. 1(c). In this grating, additional upward cavities are selected to modulate the real part and top small holes are chosen to modulate the imaginary part, similar to the sound leakage of slits [5]. The real and imaginary parts can be modulated simultaneously by combining small holes with additional cavities. Here sound-absorbing cotton is paved on the radiation holes to absorb sound energy as much as possible. Full sound absorption can be obtained with impedance matching boundary condition [40]. The relative impedance of sound-absorbing cotton used here is 1.7+0.1i around 3.1 kHz, and the absorption is about 93%. The wanted refractive index distribution can be obtained through adjusting the geometry parameters of modulators, including the height and width of additional cavities and radius of leaked small holes. The adjusting refractive index of three modulators are $n_1^1 = 0.16$, $n_1^2 = 1.03 - 0.16i$, $n_1^3 = 0.17 - 0.15i$, respectively. The modulation parameters endure a little deviation in order to make the retrieval precision meeting the fabrication precision. Three kinds of modulators are arranged side by side to form a sample piece in practical. The period of this grating is T=12 cm, and the width of each modulator is 3 cm equaling to 1/4 period. These samples were fabricated with photosensitive resin via Stereo Lithography Apparatus (SLA) 3D printing technique (0.1mm fabrication precision). In experiment, the waveguide is constructed by two paralleled rigid acrylic plates with a distance of 1.5cm. Ten sample pieces are installed periodically on the top plate to form a diffraction grating, as shown in Fig. 1(b). A loudspeaker array with 24 units is employed to generate a spatial gauss beam. The amplitude and phase of each loudspeaker can be controlled independently through an Antelope sound card with 32-channel output. The sound pressure fields inside the waveguide are scanned using a micro electro-mechanical microphone (Brüel & Kjær 4939), with a spatial resolution of 15 mm. Figure 3 presents the measured sound pressure fields at 3.1 kHz. A simulation with the same working condition as experiment configuration has been implemented for comparison. In the simulation, a pair of oblique incident gauss beams is sent from the top edge of the waveguide. The +1$^{st}$ order diffraction with negative incident angle is disappeared, while the -1$^{st}$ order with positive incident angle is still observed clearly. The size of measured areas, which are marked by white boxes in the whole waveguide, is 900 mm×360 mm. Corresponding enlarged views of simulated and measured sound pressure fields are shown in right column. The -1$^{st}$ diffraction order is much weaker than +1$^{st}$ diffraction order in experimental result, which agrees well with simulated result. The asymmetric diffraction deriving from unidirectional wave vector at EP is well demonstrated in the simulated and experimental results.

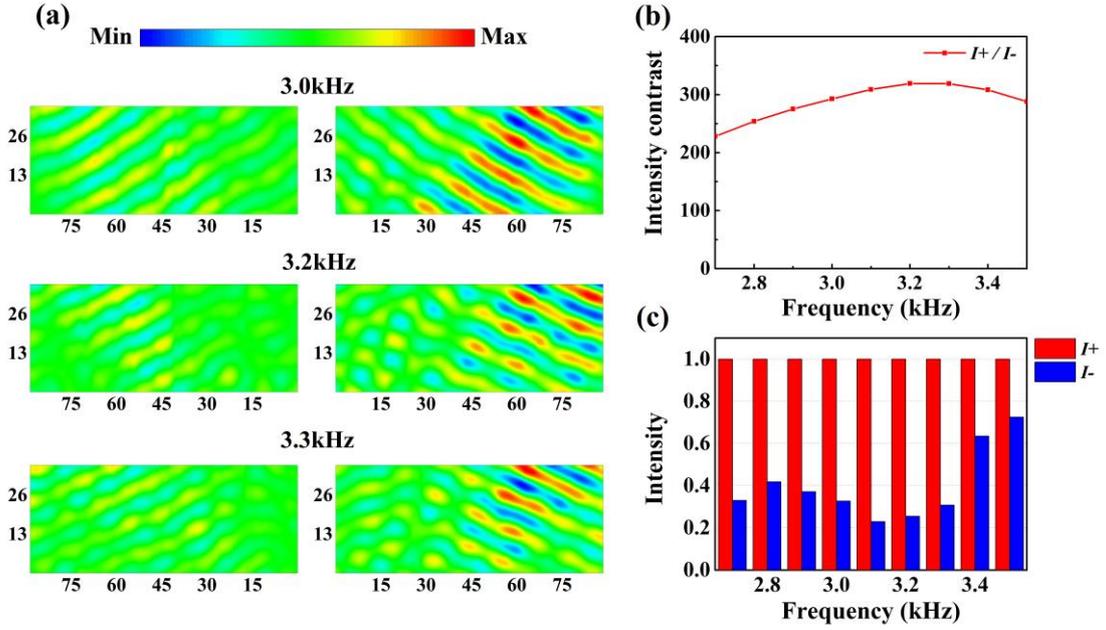

FIG. 4. Asymmetric diffraction phenomenon over a wide frequency range. (a) Measured sound pressure fields of diffracted waves at 3.0 kHz, 3.2 kHz and 3.3 kHz. The +1$^{st}$ diffraction orders are all much weaker than -1$^{st}$ diffraction orders. (b) The simulated diffraction intensity contrast under the same condition as experiment configuration. (c) Corresponding intensity bars of measured sound fields. The intensity bars of +1$^{st}$ diffraction order are all lower than those of -1$^{st}$ diffraction order.

The designed passive *PT*-symmetric grating performs asymmetric diffraction effect in a wide frequency range, since the effective refractive indexes of modulators keep steady in a wide frequency range (see Section II of Supplement Material [39]). The Bragg incident angle will change with different frequencies, as obeying the condition of $2k\sin(\theta_B) = h$. Experimental results of three other frequencies are displayed in Fig. 4(a). From the normalized sound pressure fields, it can be seen that the +1$^{st}$ diffractions with negative incident angle are much weaker than -1$^{st}$ diffractions with positive incident angle. In Fig. 4(b), we show the simulated diffraction intensity contrast I+/I- with different frequencies, which is higher than 200 in the studied frequency range. Corresponding measured results are displayed in Fig. 4(c). The intensity bars are summations of acoustic energy in the measured areas. It is obvious that the intensity bars of I- are all much lower than those of I+ in the measured frequency range. The results of experiment have demonstrated the asymmetric diffraction phenomenon of the designed *PT*-symmetric grating in a wide frequency range. Theoretically, the +1$^{st}$ diffraction order should disappear completely at EP. There are some possible reasons for weak signal of +1$^{st}$ diffraction order, including the fabrication errors of samples, incomplete absorption at the boundary of waveguides.

In conclusion, asymmetric diffraction in a passive *PT*-symmetric acoustic grating has been achieved experimentally. In this grating, the period of passive *PT*-symmetric medium is revised to eliminate the first-order disturbance in existing design. A unit combing additional cavities and leaked holes is introduced to modulate the real and imaginary parts of refractive index simultaneously. Asymmetric acoustic gratings provide a feasible way to modulate and control acoustic beams, which may have applications in noise control and beam forming. This study also broadens the route for designing functional *PT*-symmetric acoustic devices and enables us to investigate further into the physical fundamental of non-Hermitian acoustic systems.

**Acknowledgement**

This work is supported by the National Natural Science Foundation of China (Grant No. 11874383), the Youth Innovation Promotion Association CAS (Grant No. 2017029), and the IACAS Young Elite Researcher Project (Grant No. QNYC201719).